\documentclass[letterpaper]{article} % DO NOT CHANGE THIS
\usepackage[submission]{aaai24}  % DO NOT CHANGE THIS
\usepackage{times}  % DO NOT CHANGE THIS
\usepackage{helvet}  % DO NOT CHANGE THIS
\usepackage{courier}  % DO NOT CHANGE THIS
\usepackage[hyphens]{url}  % DO NOT CHANGE THIS
\usepackage{graphicx} % DO NOT CHANGE THIS
\usepackage{natbib}  % DO NOT CHANGE THIS AND DO NOT ADD ANY OPTIONS TO IT
\usepackage{caption} % DO NOT CHANGE THIS AND DO NOT ADD ANY OPTIONS TO IT
\usepackage{algorithm}
\usepackage{algorithmic}
\usepackage{url}
\usepackage{booktabs,tabularx}
\usepackage{subcaption}

\usepackage{xcolor}

\usepackage{newfloat}
\usepackage{listings}
\DeclareCaptionStyle{ruled}{labelfont=normalfont,labelsep=colon,strut=off} % DO NOT CHANGE THIS
\floatstyle{ruled}
\newfloat{listing}{tb}{lst}{}
\floatname{listing}{Listing}
\usepackage{amsfonts} 

\title{Deep Representation Learning for Open Vocabulary Electroencephalography-to-Text Decoding}
\author {
    Hamza Amrani,
    Daniela Micucci,
    Paolo Napoletano
}
\affiliations {
    University of Milano - Bicocca, Milan, Italy\\
    \{hamza.amrani, daniela.micucci, paolo.napoletano\}@unimib.it
}

\usepackage{bibentry}
\begin{document}

\maketitle

% #####################################################################################################################################
% Abstract

\begin{abstract}

Previous research has demonstrated the potential of using pre-trained language models for decoding open vocabulary Electroencephalography (EEG) signals captured through a non-invasive Brain-Computer Interface (BCI). However, the impact of embedding EEG signals in the context of language models and the effect of subjectivity, remain unexplored, leading to uncertainty about the best approach to enhance decoding performance. Additionally, current evaluation metrics used to assess decoding effectiveness are predominantly syntactic and do not provide insights into the comprehensibility of the decoded output for human understanding.
We present an end-to-end deep learning framework for non-invasive brain recordings that brings modern representational learning approaches to neuroscience. Our proposal introduces the following innovations: 1) an end-to-end deep learning architecture for open vocabulary EEG decoding, incorporating a subject-dependent representation learning module for raw EEG encoding, a BART language model, and a GPT-4 sentence refinement module; 2) a more comprehensive sentence-level evaluation metric based on the BERTScore; 3) an ablation study that analyses the contributions of each module within our proposal, providing valuable insights for future research.
We evaluate our approach on two publicly available datasets, ZuCo v1.0 and v2.0, comprising EEG recordings of 30 subjects engaged in natural reading tasks. Our model achieves  a  BLEU-1 score of 42.75\%, a ROUGE-1-F of 33.28\%,  and a BERTScore-F of 53.86\%,
outperforming the previous state-of-the-art methods by 3.38\%, 8.43\%, and 6.31\%, respectively.

%Our code is  available at: \url{https://anonymous.4open.science/r/EEG-to-Text-Decoding-E96B}

\end{abstract}

% #####################################################################################################################################
% Abstract
\section{Introduction}
\label{sec:introduction}

\begin{figure*}[tb]
\centering
\includegraphics[width=0.95\textwidth]{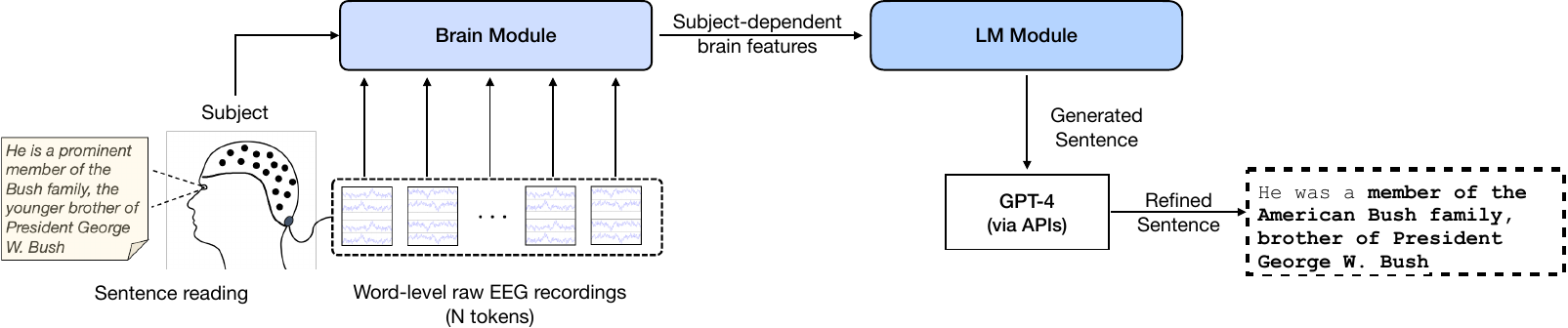} % Reduce the figure size so that it is slightly narrower than the column.
\caption{The workflow of the proposed method involves several steps. Firstly, the raw EEG signals corresponding to each word are input into the Brain module. This module extracts subject-dependent features, which are subsequently utilized by a Language Module based on BART suitable trained for sentence generation. The resulting sentence is further refined using GPT-4 APIs to produce the final output. In the example, the ground truth is: \emph{He is a prominent member of the Bush family, the younger brother of President George W. Bush}; the final sentence predicted by our model is: \emph{He was a \textbf{member of the American Bush family}, \textbf{brother of President George W. Bush}}. Bold font refers to the exact match between the ground truth and the estimated sentence.}
\label{fig1}
\end{figure*}

% introduzione a brain decoding, neuroscienze, BCIs
% ottimi risultati con approcci invasivi: parlare di closed e open vocabulary
% non invasive e open vocabulary -> challenging
% letteratura

The integration of deep learning into neuroscience is advancing rapidly. Over the past decades, Brain-Computer Interfaces (BCIs) have made significant improvements in decoding natural language from brain recordings to restore communication to people who have lost the ability to speak~\cite{willett2021high,moses2021neuroprosthesis}. Although effective, these approaches require invasive neurosurgery, making them difficult for most other uses.

Decoding methods that use non-invasive recordings could be more widely adopted, offering significant potential for application in both restorative and augmentative applications. Non-invasive brain recordings can capture multiple types of linguistic information~\cite{huth2016natural,broderick2018electrophysiological,caucheteux2022brains}, but previous attempts to use this information have been limited to decode sentences and words in small closed vocabularies~\cite{pereira2018toward, dash2020decoding, moses2021neuroprosthesis}, not clarifying whether current non-invasive recordings have the spatial and temporal resolution necessary for decoding natural language.
%leaving it unclear whether current non-invasive recordings have the spatial and temporal resolution needed to decode natural language. %aggiungere 2 citazioni una per 2022 e una per 2023
In addition, existing approaches cannot decode semantically close words.% that do not exist in the training set.

Interestingly, previous works~\cite{gauthier2018does,caucheteux2022brains} demonstrate that the human brain encodes language into higher-dimensional semantic representations.
This is similar to how modern pre-trained language models, such as BERT~\cite{devlin2018bert}, BART~\cite{lewis2019bart}, T5~\cite{raffel2020exploring}, and GPT4~\cite{OpenAI2023GPT4TR}, encode words into contextualized semantic embedded representations in Natural Language Processing (NLP). Thanks to their transfer learning abilities, diverse recent NLP downstream tasks, such as sequence classification, text generation, and question answering, have reached substantial improvements.
Likewise, various studies~\cite{wang2022open,wang2023brainbert,tang2023semantic} experimented with combining brain signal decoding to NLP models to produce semantic brain-embedded representations. % thanks to their transfer learning abilities and their efficacy in extracting semantic representations.
They demonstrate the ability of NLP models to extract semantic features that capture the meaning of input brain recordings.

%Among the state-of-the-art studies,  the one by Wang et al.~\cite{wang2022open} is quite relevant. 
The study by Wang et al.~\cite{wang2022open} is the first to prove the potential of employing pre-trained language models, such as BART, to decode open vocabulary Electroencephalography (EEG) signals captured through a non-invasive Brain-Computer Interface (BCI). The processing pipeline suggested by the authors takes as input the EEG features from the ZuCo  dataset~\cite{hollenstein2018zuco,hollenstein2019zuco}.
These pre-computed EEG features are subsequently adjusted using a transformer encoder before being input into the BART model. The BART model is then fine-tuned to effectively suit the task of decoding EEG-to-Text.
Recently, Duan et al.~\cite{duan2023dewave} present DeWave, a framework that allows for the decoding of brain dynamics into natural language without the need for eye-tracking fixations or event markers. DeWave uses a quantized variational encoder to derive discrete codex encoding and align it with a pre-trained language model. DeWave has shown superior performance compared to the state-of-the-art methods, surpassing the baseline by 3.06\% and 1.9\%, respectively, achieving 41.35\% BLEU-1 and 30.69\% Rouge-F on the ZuCo dataset.

%The achieved BLEU-1 score~\cite{papineni2002bleu} was 40.1\%, while the ROUGE-1-F score~\cite{lin2004rouge} was 30.1\%. 

Nevertheless, the unexplored impact of embedding EEG signals within language models raises questions about the optimal approach for enhancing decoding performance. Furthermore, while the analysis of EEG signals is a valuable way of studying brain activity, the interpretation of these signals can indeed be influenced by subjectivity~\cite{jeng2020low}. A recent study by Feng et al. ~\cite{feng2023semantic}, on EEG-to-Text decoding task, argued that this task is considerably challenged by the EEG representation that varies with individual subjects and the text representation influenced by semantics.  Lastly, current evaluation metrics that primarily focus on syntactic aspects do not adequately capture the semantics, resulting in limited comprehensibility.
%Lastly, the current evaluation metrics, which predominantly focus on syntactic aspects, leave much to be desired in terms of comprehensibility for human understanding.

In this paper, we present an end-to-end deep learning framework for non-invasive brain recordings that uses pre-trained language models for open vocabulary EEG-to-text decoding.
\emph{Firstly}, our end-to-end deep learning architecture for open vocabulary EEG decoding incorporates a representation learning module for raw EEG encoding, a language modeling module based on BART~\cite{lewis2019bart}, and a GPT-4~\cite{OpenAI2023GPT4TR} refinement module, enhancing the comprehensibility of the generated sentences. The representation learning module includes a subject layer,  which permits taking into account the subjectivity of EEG signals, and a multi-layer transformer encoder that allows to extract latent brain representations that are then aligned into language token embeddings. 
\emph{Second}, we use the BERTScore~\cite{zhang2019bertscore} in the evaluation, which incorporates semantic judgment at the sentence level, resulting in a more comprehensive evaluation that is closer to human perception. 
\emph{Thirdly}, we conducted an ablation study to analyze and distinguish the contributions of each module within our proposal, providing valuable insights for future research work. %(Figure~\ref{fig1}).

To demonstrate the efficacy of our approach, comprehensive evaluations are conducted on two publicly available datasets, ZuCo v1.0 and v2.0~\cite{hollenstein2018zuco,hollenstein2019zuco}, comprising EEG recordings from 30 subjects actively engaged in natural reading tasks. The results achieved by our proposal, on previously unseen sentences, are a BLEU-1 score of 42.75\%, a ROUGE-1-F~\cite{lin2004rouge} of 33.28\%, and a BERTScore-F of 53.86\%, surpassing the previous state-of-the-art results by 3.38\%, 8.43\%, and 6.31\%, respectively.

Our code is  available for public access at: \url{https://github.com/hamzaamrani/EEG-to-Text-Decoding}
\section{Related Work}

Related work on brain-to-speech and brain-to-text decoding can be categorized into three methods by the features they are capturing: \emph{motor imagery based}, \emph{overt speech based}, and \emph{inner speech based}. 

Different BCI devices have been explored encompassing Electroencephalography (EEG), Electrocorticography (ECoG), and functional Magnetic Resonance Imaging (fMRI).

%\subsection{Motor Imagery based} 
\emph{Motor imagery based} systems, such as for instance, point-and-click~\cite{jarosiewicz2015virtual, pandarinath2017high, lee2018high} and imaginary handwriting~\cite{willett2021high}, have high accuracy but moderately low typing rate. 

%\subsection{Overt Speech based} 
\emph{Overt speech based} methods for decoding or synthesizing speech show a faster communication rate. These methods require subjects to physically speak during neural recording~\cite{anumanchipalli2019speech, makin2020machine} or to imagine the physical pronunciation of the sentence~\cite{moses2021neuroprosthesis, willett2023high}.
These approaches make the decoding to be system language-dependent, since the same concept may have completely distinct pronunciations in different languages. 

%\subsection{Inner Speech based} 
\emph{Inner speech based} approaches try to address language articulation dependencies by decoding language from imagined speech and read text~\cite{brigham2010imagined, panachakel2021decoding, wang2022open, nieto2022thinking, defossez2023decoding, tang2023semantic}.

%\subsubsection{}
%Among these studies, 
A major limitation for most of the approaches discussed is the constraint of using small closed vocabularies, with a low and limited number of unique words~\cite{pereira2018toward, dash2020decoding, moses2021neuroprosthesis}.

In addition, most current approaches~\cite{willett2021high, willett2023high,defossez2023decoding} for language communication use invasive devices (such as ECoG) or less accessible non-invasive devices (such as fMRI). This makes it challenging to collect large-scale datasets and implement approaches to help people with paralysis who can no longer speak.
Nevertheless, recent studies attempt to decode inner speech by using both open vocabularies and non-invasive devices~\cite{wang2022open, defossez2023decoding, duan2023dewave}.

Our work opens the doors for similar studies of inner speech brain-to-text decoding. We investigate the representation learning of EEG signals, the inter-subject variability, the human judgment at the sentence level of generated sentences, and the use of pre-trained language models.

% #####################################################################################################################################
% Method
\section{Method} %oppure usare nome architettura/task
\label{sec:method}

%Overview of the proposed end-to-end architecture for open vocabulary EEG-to-Text decoding. Firstly, a sequence of word-level raw EEG signals is fed to the Brain module to extract deep-embedded representations for raw EEG encoding. The Brain module consists of (1) a learnable EEG feature block, (2) a subject layer to leverage inter-subject variability, (3) a multi-layer transformer (Brain Transformer Encoder), and (4) a multi-layer perceptron (MLP). Then, we use a Language Modeling module to generate and refine EEG-to-Text sentences by leveraging pre-trained language models BART and GPT-4.
\begin{figure*}[t]
\centering
\includegraphics[width=0.95\textwidth]{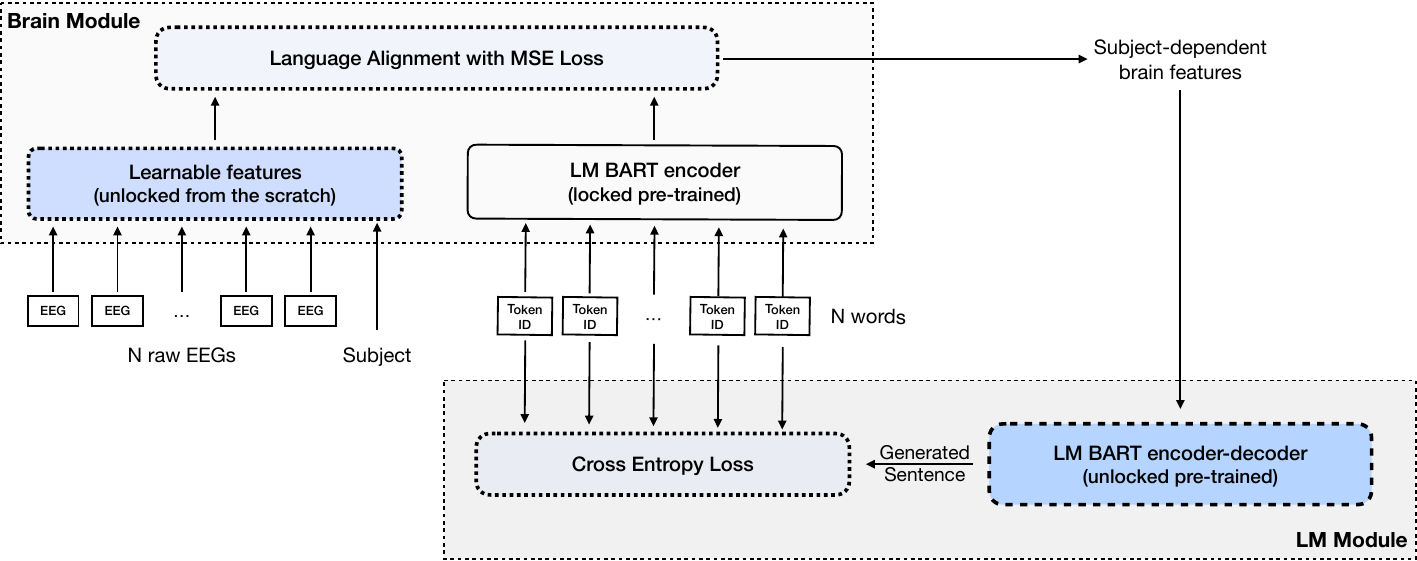} % Reduce the figure size so that it is slightly narrower than the column. Don't use precise values for figure width.This setup will avoid overfull boxes.
\caption{Overview of the proposed end-to-end architecture for open vocabulary EEG-to-Text decoding. Firstly, a sequence of word-level raw EEG signals is fed to the Brain module to extract deep-embedded representations for raw EEG encoding. Then, we use a Language Modeling (LM) module to generate EEG-to-Text sentences by leveraging the pre-trained language model BART. }
\label{fig:architecture}
\end{figure*}
% mettere immagine PDF

We aim to decode neural activity from a time series of high-dimensional brain signals recorded with non-invasive electroencephalography during the natural reading of English sentences. We first define the general task of open vocabulary EEG-to-Text decoding and then introduce the proposed end-to-end architecture. %\DAN{use va cambiato con propose}. 

\subsection{Open Vocabulary EEG-to-Text Decoding}%{Task Definition} %Neural decoding

%Given a sequence of word-level raw EEG signals $E$ and the relative subject id $s \in S$ given a group of subjects $S$, the task is to decode the corresponding text tokens from an open vocabulary $V$ in a Sequence-To-Sequence framework. In this paper, we use EEG-subject-Text pairs $<E, s, T>$ recorded in natural reading tasks, e.g., ZuCo dataset (Hollenstein et al. 2018, 2020). During the training phase, such EEG-Text pairs can come from various subjects and various categories of reading materials. During the test phase, the text sentences $T$ in $<E, s, T>$ are totally unseen.
% In this paper, we use EEG-subject-Text pairs $<X, s, Y>$ to find

%(with $|V| = 50,265$)
Let's define a sequence of word-level raw EEG signals as $X \in \mathbb{R}^{C \times T}$, with $C$ the number of EEG channels and $T$ the number of time steps. These EEG signals are a reflection of the recorded brain activity for a specific subject denoted as $s$, drawn from the set $S$ consisting of various distinct subjects. An EEG-decoding task is the task of predicting the corresponding text sentence $Y$ in a Sequence-To-Sequence framework. Each text sentence $Y$ is composed of English tokens $y_n \in \mathcal{V}$ from an open vocabulary $\mathcal{V}$. During the training phase, the EEG-subject-Text pairs can come from various subjects and various categories of reading materials. %During the test phase, the text sentences $Y$ in $<X, s, Y>$ are totally unseen.

Thus, a supervised EEG-to-Text decoding task consists in finding a decoding function $f: \{C \times T\} \times S \rightarrow \mathcal{V}$, such that $f$ predicts $Y$ given $X$ and $s$. We denote by $ \overline{Y} = f(X,s) $ the decoded/predicted text sentence from the brain signals.

Searching for $f$, the task is to maximize the probability of the decoded text sentence $\overline{Y}$:

\begin{equation}
    p(\overline{Y}|X) = \prod_{n=1}^{N} p(\overline{y}_n \in \mathcal{V} | X, \overline{y}_{<n} )
\end{equation}

where $N$ is the length of the text sentence $\overline{Y}$, and $\overline{y}_n$ is the $n$-th token of $\overline{Y}$.

\subsection{Proposed Architecture}

An overview of the proposed architecture is given in Figure~\ref{fig1} (refer to Appendix A~\ref{architecture_full} for a detailed overview of the architecture). It is composed of two main components: 1) a Brain module that implements a representation learning approach for EEG encoding; and 2) a Language Modeling module based on BART to produce EEG-to-Text sentences and on GPT-4 for sentence-level refinement.
The training process is composed of two stages. An overview of the end-to-end architecture is presented in Figure~\ref{fig:architecture}, where dashed boxes correspond to the modules of the architecture that undergo training, while solid boxes represent the module that remains untrained. We start detailing the specifics of the training stages. Then we offer a more comprehensive breakdown of each module included in our architecture.

\subsubsection{Training Stage 1} %To begin with, 
We initiate training with the Brain module: word-level EEG signals are aligned with word-tokens, as encoded by a locked, pre-trained BART Language Model, utilizing a Mean Square Error (MSE) Loss. This stage incorporates a learnable features module designed to account for EEG encoding and subjectivity. The outcome of this training stage yields EEG subject-dependent features. The alignment procedure is done by mapping the learned EEG representation $Z$ into the BART token embeddings $BART_{enc}^{te}$, using MSE regression loss $L_{MSE}( BART_{enc}^{te}, Z)$:

\begin{equation}
    \min_{f_{brain}} \mathcal{L}_{MSE}( BART_{enc}^{te}, f_{brain}(X) )
\end{equation}

\subsubsection{Training Stage 2} 
Moving on, the subsequent step involves fine-tuning a pre-trained Language Model based on BART, aimed at generating word sequences through the utilization of a Cross-Entropy Loss. As in Wang et al.~\cite{wang2022open}, we use the mapped embedded brain representation $Z$ directly as initial word embeddings to feed into the pre-trained language model encoder-decoder BART~\cite{lewis2019bart}. The high-level idea here is that we consider each embedded EEG representation as a word-level representation, and leverage a pre-trained language model to decode to real human language (English) like traditional machine translation tasks.
Then, the last hidden states from the BART decoder are fed into a multi-layer perception (MLP) to generate English tokens $\overline{y}_n$ from the BART vocabulary $\mathcal{V}$.

During the training, the objective is to minimize the text reconstruction cross-entropy loss, defined as follows:

\begin{equation}
    \mathcal{L}_{rec} = - \sum_{n=1}^{N} log \; p(\overline{y}_n \in \mathcal{V})
\end{equation}

\subsubsection{Learnable Features Module}%{Brain module}

\begin{figure}[tb]
\centering
\includegraphics[width=1.0\columnwidth]{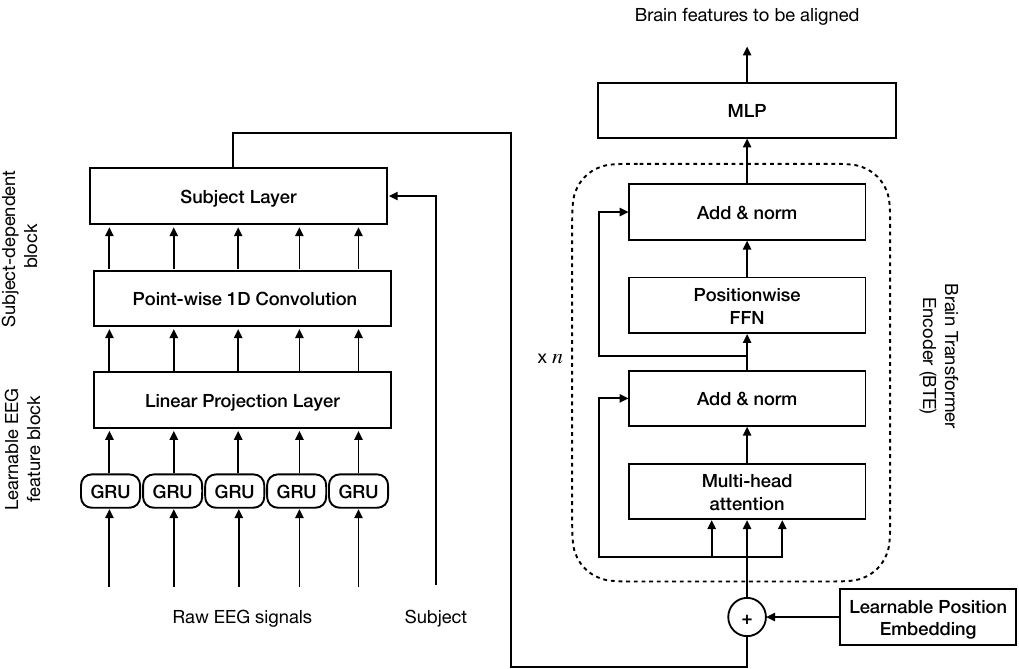}
\caption{The Learnable features module consists of (1) a learnable EEG feature block, (2) a subject layer to leverage inter-subject variability, (3) a multi-layer transformer (Brain Transformer Encoder), and (4) an MLP.}
\label{fig:brain_module}
\end{figure}

This module is included in the Brain module and it is used for extracting subject-dependent brain features from the raw EEG signals. Given a sequence of word-level raw EEG signals $X=\{x_0, x_1, ..., x_M\} \in \mathbb{R}^{C \times T}$ and the corresponding subject $s \in S$, we first use a deep neural network $f_{brain}$ to get the latent subject-dependent brain representation $Z = \{z_0, z_1, ..., z_M\} = f_{brain}(X) \in \mathbb{R}$.
This architecture (Figure~\ref{fig:brain_module}) consists of (1) a learnable EEG feature block followed (2) by a subject layer to leverage inter-subject variability, which is input to (3) a multi-layer transformer encoder named $BTE$ (Brain Transformer Encoder), and then to (4) a multi-layer perceptron.

The brain data is first fed to a bi-directional Gated Recurrent Unit (GRU)~\cite{cho2014properties} which reads the multi-time series input in both forward and backward directions to extract learnable EEG features. The use of GRU allows to dynamic address the different lengths of word-level raw EEG signals. We then apply a fully-connected layer to the concatenated forward and backward output. %rivedere
Similarly to~\cite{defossez2023decoding}, we then add a 1x1 point-wise convolution (with a kernel size of 1) without activation and a number $D$ of output channels. To leverage inter-subject variability, we learn a row vector $r_s \in \mathbb{R}^D $ for each subject $s \in S$ and apply it along the channel dimension.
We then apply a multi-layer transformer encoder~\cite{vaswani2017attention} $BTE$ with $L$ layers, each with $H$ attention heads and intermediate hidden dimension $d_h$. The inputs to the first layer $BTE_{in}^0$ are produced using a weight matrix $W_{in} \in \mathbb{R}^{d_h \times l}$ and combined with a learnable 1D position embedding $P$~\cite{dosovitskiy2020image}, which is randomly initialized. Each layer applies self-attention with causal attention masking and a feed-forward layer to the input, with layer normalization~\cite{ba2016layer} and dropout~\cite{srivastava2014dropout} being applied after. The outputs $BTE^j_{out}$ of the $j$-\emph{th} layer, become the inputs to the $(j+1)$-\emph{th} layer. % each \DAN{layer?}.  % \DAN{togliere: the inputs to}
Then, the final outputs $BTE_{out}^L$ are fed into a residual $MLP$ network, composed of two fully connected layers, obtaining the latent brain representations $z_m$. As we will demonstrate subsequently in the ablation study, opting to process the raw EEG signals using a recurrent neural network, rather than directly handling pre-computed features as performed by Wang et al.~\cite{wang2022open}, facilitates the extraction of subject-dependent nuances present in the brain recordings. These distinctive characteristics would otherwise remain entirely overlooked.

%To have better initial BART embedded representations, we decided to align the learned EEG representation $Z$ with the BART token embeddings $BART_{enc}^{te}$, using a mean squared error (MSE) regression loss $L_{MSE}( BART_{enc}^{te}, Z)$:

%\begin{equation}
%    \min_{f_{brain}} %\mathcal{L}_{MSE}( BART_{enc}^{te}, %f_{brain}(X) )
%\end{equation}

%\subsubsection{Brain encoding}%{Brain module}
%\subsubsection{Language Modeling module}

%\subsubsection{Sentence refinement}
%During the testing phase, we  propose the use of the pre-trained language model GPT-4~\cite{OpenAI2023GPT4TR} on top of the generated text sentence $\overline{Y}$. It results in significant improvements in text comprehensibility, as well as a reduction in grammatical errors and repetitive words, enhancing the utility and effectiveness of the generated text sentence. The prompt used for the refinement is as follows: 

\subsubsection{Sentence Refinement during Inference}

During the inference phase, we  propose the use of the pre-trained language model GPT-4~\cite{OpenAI2023GPT4TR} via APIs on top of the generated text sentence $\overline{Y}$. It results in significant improvements in text comprehensibility, as well as a reduction in grammatical errors and repetitive words, enhancing the utility and effectiveness of the generated text sentence. The prompt used for the refinement is as follows: 

\textit{As a text reconstructor, your task is to restore corrupted sentences to their original form while making minimum changes. You should adjust the spaces and punctuation marks as necessary. Do not introduce any additional information. If you are unable to reconstruct the text, respond with [False]. Reconstruct the following text: [text sentence $\overline{Y}$]}. %concludere? mettere prompt che abbiamo usato

% #####################################################################################################################################
% Experiment
\section{Experiments}

%\subsection{Setup} % ^ introduzione alla sezione

\subsection{Data}

We use Zurich Cognitive Language Processing Corpus (ZuCo)~\cite{hollenstein2018zuco,hollenstein2019zuco} datasets, which contain simultaneous electroencephalography and eye-tracking (ET) data recorded from natural reading tasks. The reading tasks include Normal Reading (NR) and Task-Specific Reading (TSR). The reading corpus of ZuCo are from movie reviews~\cite{socher2013recursive} and Wikipedia articles. We used data from all the subjects in ZuCo v1.0 and v2.0 (12 and 18 respectively). %\DAN{all the subjects in ZuCo v1.0 and v2.0 (12 and 18 respectively)}. %12 subjects in ZuCo v1.0 and 18 in v2.0. 
For the EEG recordings, high-density data were recorded at a sampling rate of 500 Hz with a bandpass of 0.1 to 100 Hz, using a 128-channel EEG Geodesic Hydrocel system (Electrical Geodesics). The recording reference was set at electrode Cz. We follow Hollenstein et al. steps~\cite{hollenstein2018zuco,hollenstein2019zuco} to perform data pre-processing on raw EEG signals, leading to 105 EEG channels from the scalp recordings.
 
In this paper, we use concatenated sequences of word-level raw EEG signals, which were synchronized with ET fixations.
We split each reading task’s data (by unique sentences) into train, validation, and test (80\%,10\%,10\%), as done by Wang et al.~\cite{wang2022open}. The sentences in the test set are totally unseen.
Table~\ref{table:statistics} shows the statistics of each reading task’s data.
Please refer to Appendix B~\ref{dataset_channels} for a detailed description of the electrodes used.

\begin{table}[tb]
\centering
%\resizebox{.95\columnwidth}{!}{
\caption{ZuCo datasets statistics for each reading task. NR stands for Normal Reading, while TSR stands for Task-Specific-Reading.}
\begin{tabular}{lcccc}
    \toprule
    \textbf{Reading Task} & \textbf{\#Sentences} & \textbf{\#Train} & \textbf{\#Val} & \textbf{\#Test} \\
    \midrule
    NR v1.0 & 300 & 3,609 & 467 & 456 \\
    NR v2.0 & 349 & 2,645 & 343 & 350 \\
    TSR v1.0 & 407 & 4,456 & 522 & 601 \\
    \bottomrule
\end{tabular}

\label{table:statistics}
\end{table}

%\subsection{Evaluation}

%\subsubsection{Sentence-level evaluation}

%\subsubsection{Contextual-level evaluation}

\subsection{Training Details}

% Model details
\subsubsection{Architecture Details}
For the brain module, we set the GRU layer size to 512, and the fully connected layer to 1024. The 1d convolution maps to 64 channels and the 1d subject vector size is set to 64. The BTE has 12 layers and 8 attention heads, with an intermediate hidden dimension of 4096 and GELU activations~\cite{hendrycks2016gaussian}. The last hidden states of BTE are projected on a feature space of 1024. Then, we use the large version of BART, with 12 layers for the encoder and decoder, 8 attention heads, and an intermediate hidden dimension of 4096. For GPT-4, we use OpenAI's APIs and the model version \textit{gpt-4}.

%More specifically, for the small model, we set the size of the self-attention layer, the feed-forward network, and the head to 256, 2048, and 4, respectively
\subsubsection{Optimization Settings}
During training, we use the SGD optimizer with a cyclical learning rate set with $5e-7$ and $5e-5$ as initial and upper values to update model parameters. The batch size is set to $1$ during the mapping between brain and word embeddings, and then $8$ during the training phase. The number of epochs is set to 25.
During the training phase, we freeze the brain module weights.
During inference, we use the model parameters on the best checkpoint based on the performance of the validation set.

For our architecture implementation, we use PyTorch\footnote{https://github.com/pytorch/pytorch} and Transformers (HuggingFace)\footnote{https://github.com/huggingface/transformers}  libraries.
Both Stage1 and Stage2 were trained on a workstation equipped with Ubuntu 22.04,  32GB RAM and 2 Nvidia GeForce GTX 1070 with 8GB Memory.

\subsubsection{Evaluation}
%In our experiments, we use BLEU and ROUGE metrics~\cite{papineni2002bleu, lin2004rouge} to measure the number of words shared by two sequences. However, because different words can represent the same meaning we also employ BERTScore~\cite{zhang2019bertscore}, an approach that uses machine learning to quantify whether two sequences share a meaning. %OLD
In our experiments, we use BLEU and ROUGE metrics~\cite{papineni2002bleu, lin2004rouge} to measure the number of words shared by two sequences. However, the lexical congruence may not fully encapsulate semantic similarity due to lexical variations denoting similar meanings.
To this end, we use BERTScore~\cite{zhang2019bertscore}, an approach that uses machine learning to capture the semantic similarity between two sequences by leveraging advanced language representations derived from the BERT model~\cite{devlin2018bert}. BERTScore allows the integration of semantic similarity at the sentence level, leading to a more comprehensive evaluation that aligns with human perception.

% #####################################################################################################################################
% Experiment
\section{Results}

% migliorare caption della tabella
% aggiungere: i nostri risultati in confronto alla baseline, mettiamo tutte le varianti per dimostrare i passi intermedi
% sottolineare w/o subject layer
% ablation BART ET fixations -> spiegare bene per evitare dubbi

\begin{table*}[!ht]
\centering
\caption{Open Vocabulary EEG-to-Text decoding model evaluation on ZuCo datasets. We compare our architecture (without GPT-4 sentence refinement since it is used just on the inference phase) with the current state-of-the-art by using three distinct metrics: BLEU-N ($N=1,2,3,4$), ROUGE-1 (Precision, Recall, and F1 scores), and BERTScore (Precision, Recall, and F1 scores). We also report ablations and the hypothetical upper limit for BART with fixation words when no errors are made to map EEG signals to token words. 
\textbf{Bold} numbers indicate the first best result, \underline{Underline} numbers indicate the second best result.}
\resizebox{.9\textwidth}{!}{
\begin{tabular}{l cccc ccc ccc}
    \toprule

    \textbf{Method} & \multicolumn{4}{c}{\bfseries BLEU-N (\%) $\uparrow$} & \multicolumn{3}{c}{\bfseries ROUGE-1 (\%) $\uparrow$} & \multicolumn{3}{c}{\bfseries BERTScore (\%) $\uparrow$} \\
     & N=1 & N=2 & N=3 & N=4 & R & P & F & P & R & F \\
    \midrule
    \cite{wang2022open} & 40.1 & 23.1 & 12.5 & 6.8 & 28.8 & 31.7 & 30.1 & 48.84 & 52.71 & 50.66 \\
    \cite{duan2023dewave} & 41.35 & 24.15 & 13.92 & 8.22 & 28.82 & 33.71 & 30.69 & - & - & - \\
    \midrule
    
     %\textbf{Our Architecture (before)} & \textbf{42.80} & \textbf{25.49} & \textbf{15.09} & \textbf{9.03} & \textbf{30.20} & \textbf{36.42} & \textbf{32.91} & \textbf{52.26} & \textbf{54.92} & \textbf{53.51} \\
     \textbf{Our Architecture} & \textbf{42.75} & \textbf{25.90} & \textbf{15.66} & \textbf{9.56} & \textbf{30.60} & \textbf{36.71} & \textbf{33.28} & \textbf{52.62} & \textbf{55.26} & \textbf{53.86} \\
     
     \hspace{0.75cm}w/o subject layer & \underline{41.51} & 24.41 & \underline{14.31} & 8.38 & \underline{29.22} & 35.40 & 31.92 & \underline{51.09} & \underline{53.93} & \underline{52.43} \\

    \hspace{0.75cm}w/o language alignment & 41.30 & \underline{24.50} & 14.14 & \underline{8.40} & 29.16 & \underline{35.76} & \underline{32.00} & 50.82 & 53.62 & 52.16 \\
     
     \hspace{0.75cm}w/o BTE & 35.51 & 20.51 & 12.61 & 8.98 & 25.62 & 26.38 & 25.83 & 46.44 & 50.52 & 48.34 \\
     \hspace{0.75cm}w/o BART finetuning & 28.50 & 14.35 & 7.01 & 3.38 & 21.32 & 23.07 & 22.03 & 39.67 & 47.90 & 43.13 \\
     
    \midrule
     BART with fixation words & 72.45 & 62.16 & 53.80 & 46.84 & 67.16 & 75.25 & 70.65 & 66.72 & 74.47 & 69.89 \\
    \bottomrule
\end{tabular}}
%\caption{Evaluation results on ZuCo datasets using BLEU, ROUGE, and BERT-score metrics.}

\label{table:results}
\end{table*}

%%%%
\begin{table*}[!ht]
\centering
\caption{Open Vocabulary EEG-to-Text decoding examples on ZuCo unseen test sentences. We report both predictions from our model, with and without GPT-4 sentence refinement. (1-3) are in NR v1.0, v2.0. (4) is in SR v1.0. \textbf{Bold} means exact match, \textit{Italic} indicates semantic similarity. \underline{Underline} denotes error match.}
\resizebox{.95\textwidth}{!}{
\begin{tabular}{ll|l}
    \toprule

     (1) & Ground truth & He is a prominent member of the Bush family, the younger brother of President George W. Bush... \\
     & \cite{wang2022open} & was a former \textbf{member of} the \textit{American family}, and \underline{son brother} of \textbf{President George W. Bush}... \\

     & Prediction & was the member \textbf{member} of the \textit{American family}. and \textbf{younger brother of President George W. Bush} \\
     & Prediction + GPT-4 & He was a \textbf{member of the American Bush family}, \textbf{brother of President George W. Bush}…\\
     \midrule
     
     (2) & Ground truth & Raymond Arrieta (born March 26, 1965 in San Juan, Puerto Rico) is considered by many to be one \\
     & &    of Puerto Rico’s greatest comedians. \\
     & \cite{wang2022open} & \underline{mond wasaga},19 in 17, 18) \underline{New Francisco}, \textbf{Puerto Rico}) is a one many to be the of the \textit{Rico’s} \textbf{greatest}\\
     & &  \underline{poets}. \\

     & Prediction & \underline{mond wasaga} \textbf{(born} April 17, 1946) \underline{New Francisco}, \textbf{Puerto Rico}) is a one many to be the of the \textit{Rico’s} \textbf{most} \textit{artists}. \\     
     & Prediction + GPT-4 & \underline{Ramon Wasaga} \textbf{(born} April 17, 1946, in \underline{New Francisco}, \textbf{Puerto Rico})  is one of the many to be considered \\
     & & as one of \textbf{the \textit{most prominent artists} of Puerto Rico}.\\
     \midrule

    (3) & Ground truth & Following the 1980 presidential election, Bush and his family moved to Miami-Dade County, Florida. \\
    & \cite{wang2022open} &  the \underline{deaths} \textbf{election}, the was \textit{his wife} moved to \underline{California},\textbf{Dade County, Florida}\\

    & Prediction & the \underline{wars} \textbf{election}, \textbf{Bush} was \textit{his wife} moved to Florida,\textbf{Dade County, Florida}.\\
    & Prediction + GPT-4 & \textit{After} the \underline{war's} \textbf{election}, \textbf{Bush} and \textit{his wife} moved to \textbf{Dade County, Florida}.\\
    \midrule
        
     (4) & Ground truth &  It’s not a particularly good film, but neither is it a monsterous one. \\
     & \cite{wang2022open} &  was a a \textit{bad} \textbf{good} \underline{story}, but it is it \textit{bad bad.} one. \\
     
     & Prediction & 's a a \textit{bad} \textbf{good movie}, but it is it \textit{bad bad.} one.\\
     & Prediction + GPT-4 & It's a \textit{bad} \textbf{good movie}, but is it a \textit{bad} one.\\

    \bottomrule
\end{tabular}}

\label{table:results_examples}
\end{table*}
%%%

%\subsection{Proposed method evaluation}
\subsection{Improving Decoding Accuracy}

%Table~\ref{table:results} shows the results of the proposed architecture in comparison with the state-of-the-art. Our proposal achieves remarkable results, with a 42.75\% BLEU-1 score, a ROUGE-1-F of 33.28\%,  and a 53.86\% BERTScore-F, demonstrating an improvement, on average, over the state-of-the-art.% by approximately 3\%.

We compared our architecture with the current state-of-the-art models by Wang et al.~\cite{wang2022open} and Duan et al.~\cite{duan2023dewave}. As shown in Table~\ref{table:results}, our proposal achieves a BLEU-1 score of 42.75\%, a ROUGE-1-F of 33.28\%, and a BERTScore-F of 53.86\%, showing an improvement over the state-of-the-art by 3.38\%, 8.43\%, and 6.31\%, respectively. For larger $n$-grams evaluation, we obtain BLEU-\{2,3,4\} scores of 25.90\%, 15.66\%, and 9.56\% respectively, leading to an increase of 7.24\%, 12.5\%, and 16.30\%. Our decoding embeddings resulted in higher performance for each metric, demonstrating the positive impact of learning embedded EEG representations and exploiting intersubject variability.
In Appendix C~\ref{results_by_subjects} we report the obtained results of our architecture for each subject. The results show a significant difference between v1.0 and v2.0 participants. On average, v2.0 participants outperform v1.0 participants by 19.64\%, 42.61\%, and 11.83\% for BLEU-1, ROUGE1-F, and BERTScore-F respectively.

In addition to numerical results, we report decoding examples of generated EEG-to-Text sentences compared to the ground truth and the state of the art, with and without GPT-4 sentence refinement (Table~\ref{table:results_examples}). We observe that our model is sometimes able to precisely capture named entities that do not exist in the training set. ``George W. Bush'' in (1) and ``Puerto Rico'' in (2) are correctly decoded, while ``presidential election'' in (3) is incorrectly decoded.
Compared to~\cite{wang2022open}, our model results in significant improvements in text comprehensibility, as well as a reduction in grammatical errors and repetitive words, as shown in example (4). Please refer to Appendix D~\ref{decoding_examples_appendix} to see additional decoding examples of generated EEG-to-Text sentences.

%Although the observed improvements in the performance of our model are modest, it is important to contextualize these advances within the challenging nature of open vocabulary EEG decoding and interpretation of non-invasive brain recordings. 
The complexity of open vocabulary EEG decoding tasks arises from the high dimensionality, intersubjectivity, and variability of EEG data, coupled with the intrinsic difficulties associated with the language decoding capabilities of AI-based language models.
Our improvements represent significant progress in overcoming these multiple challenges and suggest a promising direction for future research in non-invasive brain decoding.

\subsubsection{Embedding Visualization}
\begin{figure*}[tb]
    \centering
    \includegraphics[width=0.4\textwidth]{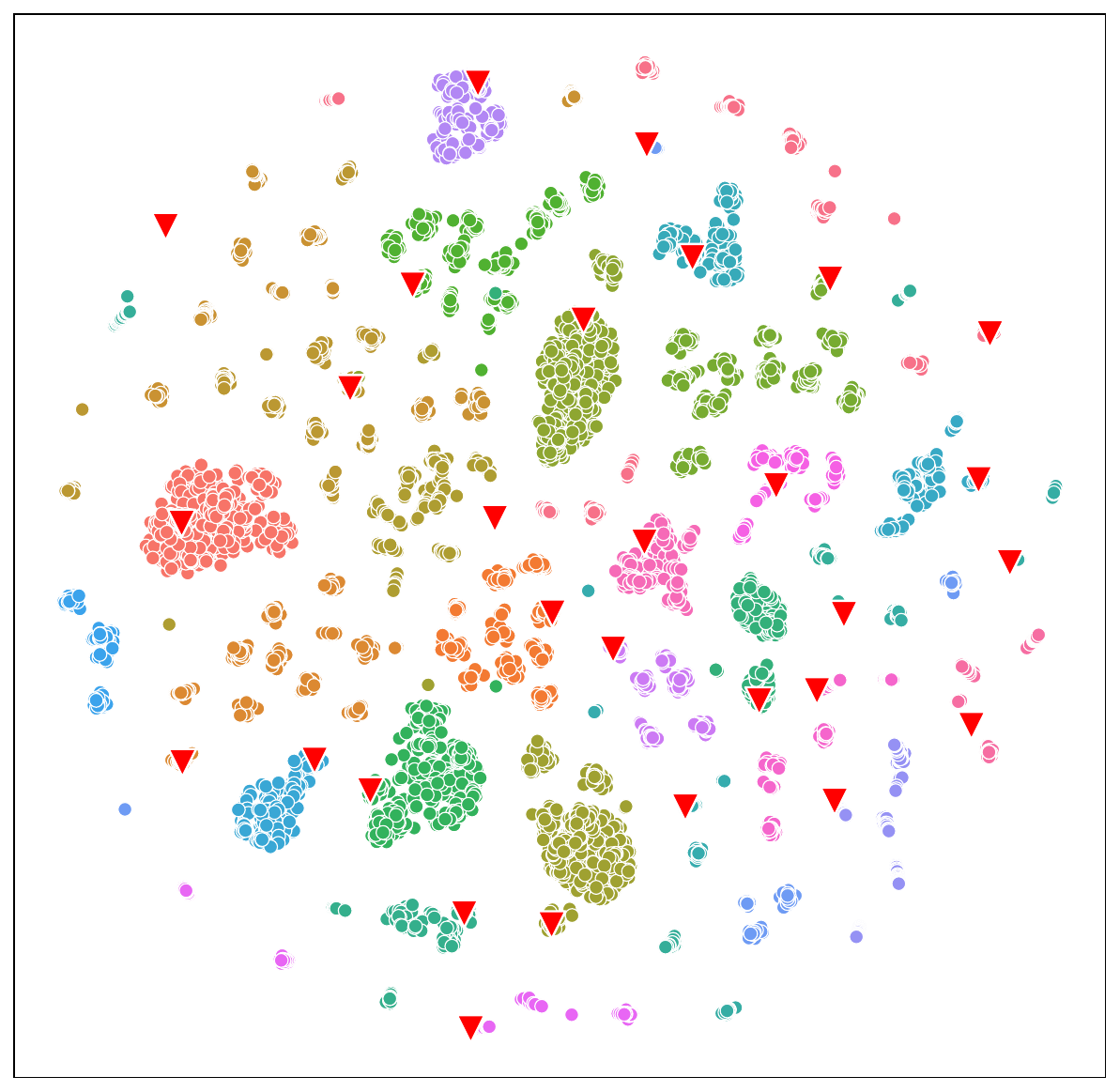}
    \includegraphics[width=0.47\textwidth]{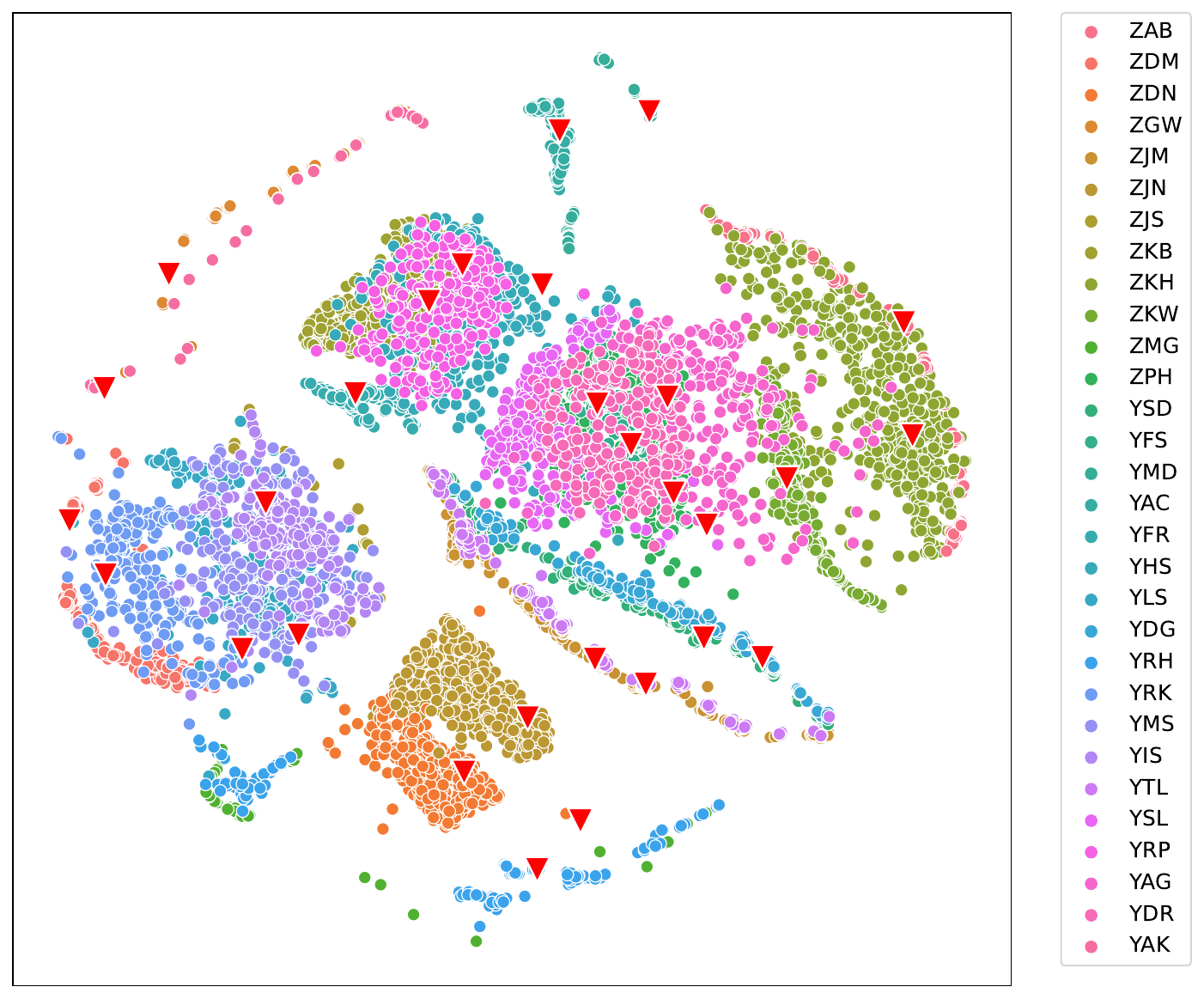}
    \caption{t-SNE visualization of EEG embedded representations of sentences in the training set, which are (a) original EEG representations and (b) generated by the Brain module of our architecture. Distinct colors mean different subjects. Each dot represents a sentence. The red triangle represents the EEG embedded representations corresponding to the same sentence \textit{"With his interest in race cars, he formed a second company, the Henry Ford Company".} }
    \label{fig:embeddings}
\end{figure*}

%\subsection{Representations Visualization}
We provide a visual comparison via t-distributed stochastic neighbor embedding (t-SNE) between the precalculated EEG features (Figure~\ref{fig:embeddings} (left)) as used by Wang et al.~\cite{wang2022open}, and EEG embedded representations obtained by the proposed Brain module (Figure~\ref{fig:embeddings} (right)). Distinct colors refer to different subjects. Each dot represents a sentence. The red triangle represents the EEG embedded representations corresponding to the same sentence \textit{``With his interest in race cars, he formed a second company, the Henry Ford Company.''} We can observe that our learned EEG representations of sentences from the same subject are much more grouped compared with pre-calculated EEG representations, denoting the capacity of our latent space to model EEG subjectivity. %. However, we can notice that, in some cases,  learned EEG representations of the same sentence are not fully clustered. See for example, Instead, multiple subject clusters are constructed indicating high inter-subject variability. This also indicates that achieving sentence-dependent EEG embedded representations is a challenging task.

%\subsubsection{Subject variability}

\subsection{Ablations}

Our ablations highlight the importance of (1) the subject layer, (2) the language alignment, (3) the use of the Brain Transformer Encoder, and (4) the BART finetuning (Table~\ref{table:results}). 
First, a model trained to generate EEG-to-Text sentences without the use of the subject layer achieves lower decoding accuracy on average across datasets, that is, about $1-1.5\%$ lower than our model. While modest, these scores show the positive effect of leveraging inter-subject variability.
Second, we show the effect of using language alignment with MSE. The results show small differences, especially in the BLEU and ROUGE scores. For BERTScore we see small improvements.
Thirdly, sentence generation without the Brain Transformer Encoder shows a significant drop in performance compared to our model. 
For BLEU-1 the decrease is $7.24\%$, while for BLEU-2 is $5.39\%$. While, ROUGE-1-F and BERTScore-F lose $7.45\%$ and $5.52\%$, respectively. We verified that the Brain Transformer Encoder provides higher decoding performances.
Finally, to test whether our model effectively leverages the pre trained BART model, we trained it without fine-tuning the BART model weights. As reported, decoding performance decreases notably up to $14.25\%$. This loss significantly confirms the use of fine-tuning on the BART model.

Then, we also show the hypothetical upper limit for EEG-to-Text decoding when no errors are made to map EEG signals to token words. Separately from our model, we fine-tuned BART on only Eye-Tracking fixations words without considering the raw EEG signals to reconstruct the original text sentence. It outperforms our proposed architecture by about 30\% in terms of BLEU-1, 37\% in terms of ROUGE-1-F, and 15\% in terms of BERTScore-F. 
The obtained results reveal the existence of two challenges within the EEG-to-Text decoding task. The initial challenge pertains to the model's capacity to establish a dependable EEG-feature representation for the word tokens. The subsequent challenge involves the faithful reconstruction of the sentence. This experiment highlights that, between these two challenges, the foremost one is undoubtedly the ability to discern an efficacious representation of the EEG signals. This observation thereby points towards the direction of future research efforts.

%\subsubsection{Results without the subject layer}
%\subsubsection{Results without the Brain Transform Encoder}
%\subsubsection{Results without BART finetuning}

% 1. training senza pezzi
% 2. upper limit
% 3. scelta di BART e non altri modelli
% 4. prova di training di allineamento con encoder

\subsection{Ethical Implications} %Ethical considerations and societal impact
While the recent advancements in utilizing brain-computer interfaces and artificial intelligence to decode neural activity into text hold significant potential in aiding individuals with communication deficits, ethical considerations, and societal impact must be carefully addressed. The scientific community must maintain vigilance and ensure that the utilization of such systems is not employed without the informed and declared consent of the participants. Fortunately, the current nature of acquiring EEG and MEG (Magnetoencephalography) signals requires participant awareness, unlike other biomarkers such as DNA or facial features. Additionally, the susceptibility of these signals to corruption by muscle movements, such as teeth clenching or eye blinks, provides a possible precaution against unauthorized acquisition and misuse. 
Furthermore, it is critical to acknowledge the potential risk associated with the high subjectivity of neural signals, even in the absence of participant awareness, which could compromise mental privacy.

We strongly believe that promoting and encouraging open science practices remains essential for responsibly assessing the potential risks and benefits associated with BCI and AI technologies in this domain.

% 

% #####################################################################################################################################
% Conclusion
\section{Conclusions and Future Works} %Discussion

In this paper, we present an end-to-end deep learning framework for open vocabulary EEG-to-Text decoding task. % by leveraging subject-dependent EEG representation features and pre-trained language models. 
By leveraging a subject-dependent representation learning module, a pre-trained BART language model, and a GPT-4 sentence refinement module, this study offers a comprehensive solution that not only enhances decoding performance but also delves into the human comprehensibility of the decoded output. The incorporation of the BERTScore as an evaluation metric has enabled a more holistic assessment, capturing not only syntactic accuracy but also taking into account human understanding at the sentence level. Moreover, the conducted ablation study permitted us to understand the contribution to the proposed architecture of each component. This in-depth analysis not only validates the efficacy of each module but also provides a roadmap for further research, guiding the development of refined and optimized approaches in the future.

The empirical validation on two publicly available datasets demonstrates the effectiveness of the proposed architecture, achieving a BLEU-1 score of 42.75\%, a ROUGE-1-F of 33.28\%, and a BERTScore-F of 53.86\%, outperforming the previous state-of-the-art results by 3.38\%, 8.43\%, and 6.31\%, respectively. When looking at larger $n$-grams ratings (BLEU-{2,3,4}), there is an improvement of 7.24\%, 12.5\%, and 16.30\%, respectively.
Our results show that the use of raw EEG signals leads to improved results, demonstrating the effectiveness of modern representational learning approaches in neuroscience.

In summary, this research not only fills critical voids in the EEG decoding landscape but also shows the way for future investigations. By combining advanced neural network architectures with sophisticated evaluation methodologies, the study pushes the boundaries of EEG-to-text decoding and encourages continued innovation in the pursuit of more accurate and human-aligned results.

One future direction is to improve the quality of the generated embedded representations by taking into account inter-subject variability, so to increase the ability of the model to generalize across individuals.
Furthermore, ethical considerations need to be at the forefront as we move forward. Ensuring privacy, establishing clear guidelines for consent, and considering the potential long-term effects of this technology on users are critical.

%We present an end-to-end architecture for learning embedded representations of brain activity during natural reading with non-invasive brain recordings. The resulting representations have three main benefits. 
%First, they improve the accuracy efficiency of open vocabulary EEG-to-Text decoding. Compared to the baseline, there is an improvement of 6.60\% for BLEU-1, 10.56\% for ROUGE-1, and 6.31\% for BERTScore.
% commento confronto con risultati
%Second, they provide this performance improvement by being simultaneously trained across several participant with a designed subject layer to leverage inter-subject variability.
%commento sull'incremento delle performance migliorate e plot visivo
%Finally, by using pre-trained language model BART, they capture syntactic features, semantic features, and long-distant dependencies of natural language.
%commento su BART. perchè va bene con embedded eeg

%To further refine the embedded representations that our model generates, in the future, we intend to i) extend training with additional brain recordings from numerous subjects, and ii) focus on subject variability and personalization.

% commentare per review
\section*{Acknowledgement}

%\textbf{da commentare prima della submission }
%Omitted for double blind revision. 

This work was partially funded by the National Plan for NRRP Complementary Investments (PNC, established with the decree-law 6 May 2021, n. 59, converted by law n. 101 of 2021) in the call for the funding of research initiatives for technologies and innovative trajectories in the health and care sectors (Directorial Decree n. 931 of 06-06-2022) - project n. PNC0000003 - AdvaNced Technologies for Human-centrEd Medicine  (project acronym: ANTHEM). This work reflects only the authors’ views and opinions, neither the Ministry for University and Research nor the European Commission can be considered responsible for them.

%t #####################################################################################################################################
% After main content
%\appendix

%\section{Appendices}

%\section{Ethical Statement}

%\section{Acknowledgments}

% #####################################################################################################################################
% Bibliography
\bibliography{aaai24}

%t #####################################################################################################################################
\appendix

\clearpage

\onecolumn 
\section{Appendix}

\subsection{A - Architecture}
\label{architecture_full}

\begin{figure*}[h]
\centering
\includegraphics[width=0.95\textwidth]{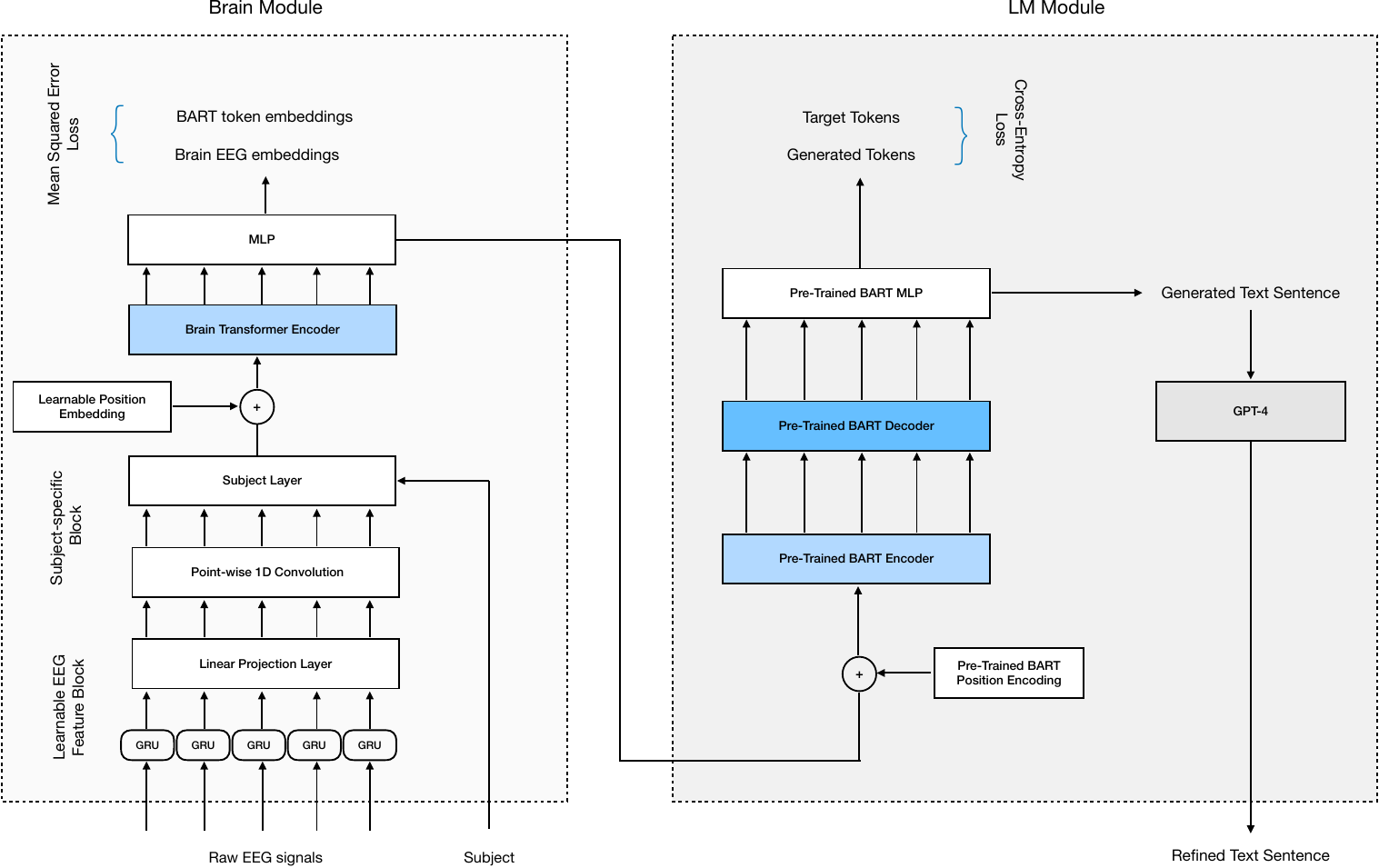}
\caption{End-to-end architecture for open vocabulary EEG-to-Text decoding.}
\label{fig:architecture_full}
\end{figure*}

A detailed overview of the architecture is given in Figure~\ref{fig:architecture_full}. It is composed of two main components: 1) a Brain module that implements a representation learning approach for EEG encoding; and 2) a Language Modeling module based on BART to produce EEG-to-Text sentences and on GPT-4 for sentence-level refinement.

\subsection{B - Dataset EEG electrodes}%{EEG Electrodes}
\label{dataset_channels}
In ZuCo dataset~\cite{hollenstein2018zuco,hollenstein2019zuco}, we follow Hollenstein et al. steps~\cite{hollenstein2018zuco,hollenstein2019zuco} to perform data pre-processing on raw EEG signals, leading to 105 EEG channels from the scalp recordings.
It follows the full list of EEG channels: \textit{E2, E3, E4, E5, E6, E7, E9, E10, E11, E12, E13, E15, E16, E18, E19, E20, E22, E23, E24, E26, E27, E28, E29, E30, E31, E33, E34, E35, E36, E37, E38, E39, E40, E41, E42, E43, E44, E45, E46, E47, E50, E51, E52, E53, E54, E55, E57, E58, E59, E60, E61, E62, E64, E65, E66, E67, E69, E70, E71, E72, E74, E75, E76, E77, E78, E79, E80, E82, E83, E84, E85, E86, E87, E89, E90, E91, E92, E93, E95, E96, E97, E98, E100, E101, E102, E103, E104, E105, E106, E108, E109, E110, E111, E112, E114, E115, E116, E117, E118, E120, E121, E122, E123, E124, Cz}.

In this paper, the Cz EEG channel has been removed as it consists of all zeros. 

\newpage
\subsection{C - Decoding accuracy results by subject}
\label{results_by_subjects}

We report open vocabulary EEG-to-Text decoding results for each subject (see Table~\ref{table:results_by_subject}).
The results show a significant difference between subjects from the v1.0 and v2.0 of the dataset. 
The v2.0 results achieve a BLEU-1 score of 47.13\%, a ROUGE-1-F of 40.16\%, and a BERTScore-F of 57.35\%, while the v1.0 results obtain a BLEU-1 score of 39.39\%, a ROUGE-1-F of 28.16\%, and a BERTScore-F of 51.28\%, so leading to an increment of 19.64\%, 42.61\% and 11.83\% respectively.

\begin{table*}[!ht]
\centering
\caption{Open Vocabulary EEG-to-Text decoding model evaluation on ZuCo datasets by each subject.} 
\resizebox{.9\textwidth}{!}{
\begin{tabular}{l c cccc ccc ccc}
    \toprule

    \textbf{Subject} & \textbf{ZuCo} & \multicolumn{4}{c}{\bfseries BLEU-N (\%) $\uparrow$} & \multicolumn{3}{c}{\bfseries ROUGE-1 (\%) $\uparrow$} & \multicolumn{3}{c}{\bfseries BERTScore (\%) $\uparrow$} \\
     & & N=1 & N=2 & N=3 & N=4 & R & P & F & P & R & F \\
    \midrule

    ZAB & v1.0 & 39.38 & 22.11 & 11.92 & 6.61 & 25.94 & 30.92 & 28.11 & 49.88 & 52.62 & 51.16 \\
    ZDM & v1.0 & 39.45 & 22.24 & 12.02 & 6.67 & 25.93 & 30.94 & 28.11 & 50.00 & 52.73 & 51.28 \\
    ZDN & v1.0 & 39.06 & 21.93 & 11.81 & 6.63 & 26.12 & 31.25 & 28.35 & 49.80 & 52.45 & 51.04 \\
    ZGW & v1.0 & 39.79 & 22.57 & 12.27 & 6.92 & 26.08 & 30.98 & 28.22 & 50.34 & 53.07 & 51.62 \\
    ZJM & v1.0 & 39.27 & 21.99 & 11.97 & 6.67 & 25.94 & 30.96 & 28.12 & 49.73 & 52.46 & 51.00 \\
    ZJN & v1.0 & 39.76 & 22.52 & 12.49 & 7.05 & 26.51 & 31.48 & 28.68 & 50.37 & 53.07 & 51.64 \\
    ZJS & v1.0 & 39.22 & 22.49 & 12.23 & 6.82 & 25.66 & 30.29 & 27.69 & 50.47 & 53.23 & 51.76 \\
    ZKB & v1.0 & 39.38 & 22.11 & 11.92 & 6.61 & 25.94 & 30.92 & 28.11 & 49.88 & 52.62 & 51.16 \\
    ZKH & v1.0 & 39.32 & 22.01 & 11.86 & 6.60 & 26.00 & 31.00 & 28.18 & 49.86 & 52.60 & 51.14 \\
    ZKW & v1.0 & 39.38 & 22.11 & 11.92 & 6.61 & 25.94 & 30.92 & 28.11 & 49.88 & 52.62 & 51.16 \\
    ZMG & v1.0 & 39.29 & 22.22 & 12.02 & 6.70 & 25.95 & 30.93 & 28.12 & 49.93 & 52.68 & 51.22 \\
    ZPH & v1.0 & 39.38 & 22.11 & 11.92 & 6.61 & 25.94 & 30.92 & 28.11 & 49.88 & 52.62 & 51.16 \\
    YSD & v2.0 & 47.22 & 30.64 & 20.04 & 12.94 & 36.80 & 44.47 & 40.19 & 56.10 & 58.63 & 57.29 \\
    YFS & v2.0 & 47.09 & 30.87 & 20.29 & 13.15 & 36.76 & 44.65 & 40.24 & 56.21 & 58.68 & 57.37 \\
    YMD & v2.0 & 47.22 & 30.64 & 20.04 & 12.94 & 36.80 & 44.47 & 40.19 & 56.10 & 58.63 & 57.29 \\
    YAC & v2.0 & 46.88 & 30.25 & 19.92 & 12.90 & 36.59 & 44.62 & 40.14 & 56.22 & 58.52 & 57.30 \\
    YFR & v2.0 & 45.82 & 29.23 & 19.09 & 12.21 & 35.91 & 42.64 & 38.91 & 56.51 & 59.13 & 57.74 \\
    YHS & v2.0 & 47.22 & 30.55 & 20.00 & 12.92 & 36.80 & 44.41 & 40.16 & 56.06 & 58.60 & 57.26 \\
    YLS & v2.0 & 47.22 & 30.64 & 20.04 & 12.94 & 36.80 & 44.47 & 40.19 & 56.10 & 58.63 & 57.29 \\
    YDG & v2.0 & 47.22 & 30.64 & 20.04 & 12.94 & 36.80 & 44.47 & 40.19 & 56.10 & 58.63 & 57.29 \\
    YRH & v2.0 & 47.22 & 30.64 & 20.04 & 12.94 & 36.80 & 44.47 & 40.19 & 56.10 & 58.63 & 57.29 \\
    YRK & v2.0 & 47.22 & 30.64 & 20.04 & 12.94 & 36.80 & 44.47 & 40.19 & 56.10 & 58.63 & 57.29 \\
    YMS & v2.0 & 47.22 & 30.64 & 20.04 & 12.94 & 36.80 & 44.47 & 40.19 & 56.10 & 58.63 & 57.29 \\
    YIS & v2.0 & 47.22 & 30.64 & 20.04 & 12.94 & 36.80 & 44.47 & 40.19 & 56.10 & 58.63 & 57.29 \\
    YTL & v2.0 & 47.22 & 30.64 & 20.04 & 12.94 & 36.80 & 44.47 & 40.19 & 56.10 & 58.63 & 57.29 \\
    YSL & v2.0 & 47.52 & 31.00 & 20.34 & 13.20 & 37.23 & 44.98 & 40.65 & 56.54 & 59.02 & 57.71 \\
    YRP & v2.0 & 47.22 & 30.64 & 20.04 & 12.94 & 36.80 & 44.47 & 40.19 & 56.10 & 58.63 & 57.29 \\
    YAG & v2.0 & 47.22 & 30.64 & 20.04 & 12.94 & 36.80 & 44.47 & 40.19 & 56.10 & 58.63 & 57.29 \\
    YDR & v2.0 & 47.16 & 30.63 & 20.23 & 13.17 & 37.00 & 44.70 & 40.40 & 56.31 & 58.74 & 57.45 \\
    YAK & v2.0 & 47.22 & 30.64 & 20.04 & 12.94 & 36.80 & 44.47 & 40.19 & 56.10 & 58.63 & 57.29 \\

    \midrule
    
    \textbf{Average} & v1.0 & 39.39 & 22.2 & 12.03 & 6.71 & 26.0 & 30.96 & 28.16 & 50.0 & 52.73 & 51.28 \\
     & v2.0 & 47.13&  30.57&  20.02&  12.93&  36.77&  44.42&  40.16&  56.17&  58.68&  57.35 \\
      & v1.0 + v2.0 & 42.75 & 25.90 & 15.66 & 9.56 & 30.60 & 36.71 & 33.28 & 52.62 & 55.26 & 53.86 \\

    \bottomrule
\end{tabular}}

\label{table:results_by_subject}
\end{table*}

\newpage
\subsection{D - Decoding Examples}
\label{decoding_examples_appendix}
We report additional decoding examples of generated EEG-to-Text sentences (see Table~\ref{table:results_examples_appendix}), with and without GPT-4 sentence refinement.
The prompt used for the GPT-4 sentence refinement is as follows:  

\textit{As a text reconstructor, your task is to restore corrupted sentences to their original form while making minimum changes. You should adjust the spaces and punctuation marks as necessary. Do not introduce any additional information. If you are unable to reconstruct the text, respond with [False]. Reconstruct the following text: [text sentence $\overline{Y}$]}.

\begin{table*}[!ht]
\centering
\caption{Open Vocabulary EEG-to-Text decoding examples on ZuCo unseen test sentences, with and without GPT-4 sentence refinement.}
\resizebox{.95\textwidth}{!}{
\begin{tabular}{ll|l}
    \toprule

     (1) & Ground truth & An amateurish, quasi-improvised acting exercise shot on ugly digital video.\\
     & Prediction &  interesting actor, un-religiousprovised film performance, through a, video.\\
     & Prediction + GPT-4 & Interesting actor, un-religious, improvised film performance, through a video.\\
     \midrule
     (2) & Ground truth & Viewed as a comedy, a romance, a fairy tale, or a drama, there's nothing remotely  \\
     & & triumphant about this motion picture.\\
     & Prediction & the from a kind of it satire, and love tale, and a love, it's a quite funny about it film picture. \\
     & Prediction + GPT-4 & From a kind of satire, it's a love tale and quite a funny film picture about love. \\
     \midrule
     (3) & Ground truth & It's solid and affecting and exactly as thought-provoking as it should be.\\
     & Prediction &  's a, well. it what it-provoking as the sounds be.\\
     & Prediction + GPT-4 & Well, it's as provoking as it sounds, what a be.\\
     \midrule
     (4) & Ground truth & It's a head-turner -- thoughtfully written, beautifully read and, finally, deeply humanizing.\\
     & Prediction & s a greatyscing, a to crafted, well acted, well most, a moving.. \\
     & Prediction + GPT-4 & It's a great, most moving, well-crafted and well-acted scene. \\
     \midrule
     
     (5) & Ground truth & ``The Kid Stays in the Picture'' is a great story, terrifically told by the man who wrote \\
     & & it but this Cliff Notes edition is a cheat.\\
     & Prediction & The movie''ays in the House'' is a film movie about andally funny by a young \\
     & &who wrote it. also ish version is a little. \\
     & Prediction + GPT-4 & "The movie, 'Days in the House', is a film about a young man who wrote it. \\
     & &It's also randomly funny. The British version is a little different.\\
     \midrule
     
     (6) & Ground truth & Fans of the TV series will be disappointed, and everyone else will be slightly bored.\\
     & Prediction & of the film series will recognize familiar to but the will will be happy disappointed. \\
     & Prediction + GPT-4 & of the film series will recognize familiar to but the will be happy disappointed.\\
     \midrule
     (7 ) & Ground truth & Wedding feels a bit anachronistic\\
     & Prediction &  alting bells like little likeachronistic, \\
     & Prediction + GPT-4 & alting bells like little likeachronistic.\\
     \midrule
     (8) & Ground truth & But what's nice is that there's a casual intelligence that permeates the script.\\
     & Prediction &  he's most about that it's a sense, to'sates the film. \\
     & Prediction + GPT-4 & He's most about that. It's a sense to states the film.\\
     \midrule
     (9) & Ground truth & An important movie, a reminder of the power of film to move us and to make us examine our values.\\
     & Prediction &  interesting part about but must of the importance of the to shape people. of make us think our lives. \\
     & Prediction + GPT-4 & interesting part about but must of the importance of the to shape people. of make us think our lives.\\
     \midrule
     (10) & Ground truth & Jeb Bush was born in Midland, Texas, where his father was running an oil drilling company.\\
     & Prediction &  uan Bush was born in Newland, Texas, and his father was a a insurance company company. \\
     & Prediction + GPT-4 & Juan Bush was born in Newland, Texas, and his father was an insurance company owner.\\

    \bottomrule
\end{tabular}}

\label{table:results_examples_appendix}
\end{table*}

\end{document}